# Automatic finite element implementation of hyperelastic material with a double numerical differentiation algorithm

**Wang, Yuxiang**[1]
Department of Systems and Information Engineering, University of Virginia
Department of Mechanical and Aerospace Engineering, University of Virginia
151 Engineers Way, Olsson Hall, Charlottesville, VA 22903
yw5aj@virginia.edu

**Gerling, Gregory J.**
Department of Systems and Information Engineering, University of Virginia
Department of Biomedical Engineering, University of Virginia
151 Engineers Way, Olsson Hall, Charlottesville, VA 22903
gg7h@virginia.edu

**ABSTRACT**

*In order to accelerate implementation of hyperelastic materials for finite element analysis, we developed an automatic numerical algorithm that only requires the strain energy function. This saves the effort on analytical derivation and coding of stress and tangent modulus, which is time-consuming and prone to human errors. Using the one-sided Newton difference quotients, the proposed algorithm first perturbs deformation gradients and calculate the difference on strain energy to approximate stress. Then, we perturb again to get difference in stress to approximate tangent modulus. Accuracy of the approximations were evaluated across the perturbation parameter space, where we find the optimal amount of perturbation being $10^{-6}$ to obtain stress and $10^{-4}$ to obtain tangent modulus. Single element verification in ABAQUS with Neo-Hookean material resulted in a small stress error of only $7 \times 10^{-5}$ on average across uniaxial compression and tension, biaxial tension and simple shear situations. A full 3D model with Holzapfel anisotropic material for artery inflation generated a small relative error of $4 \times 10^{-6}$ for*

---

[1] Corresponding author.

*inflated radius at 25 kPa pressure. Results of the verification tests suggest that the proposed numerical method has good accuracy and convergence performance, therefore a good material implementation algorithm in small scale models and a useful debugging tool for large scale models.*

**INTRODUCTION**

Finite element (FE) analysis for biological tissues is a fundamental tool in biomechanical engineering. Often, such soft materials undergo large deformations beyond the linear range [1]. In these cases, hyperelastic materials should be used to guarantee accuracy and convergence of numerical modeling. Fully defined by their strain energy functions, various hyperelastic models were developed for different tissues. For example, the Holzapfel model depicts the behavior of artery walls [2], Fung model are often used for heart valves [3,4], and Ogden model is widely used for skin [5,6]. Numerous new models are also being actively developed, and their implementation heavily relies on the user defined material subroutines for commercial FE packages like ABAQUS [7] or FEBio [8].

Nevertheless, numerical implementation of hyperelastic material for FE analysis is a painstaking task that requires tremendous effort. The first step is to derive the explicit form of stress tensor by differentiating the strain energy function with respect to the strain tensor, and then obtain the tangent modulus tensor by differentiating the stress tensor again with respect to the strain tensor. Because strain energy functions appear in various forms and often only implicitly tied to the strain tensor - for example, Mooney-Rivlin model is expressed in strain invariants and Ogden model in principal stretches - the differentiation needs to be expanded with chain rules in a fourth order tensor space. To appreciate the details about these derivations, readers might refer to existing literature [9–11], and will immediately notice that these require non-trivial familiarity with tensor algebra. In addition, the long resultant analytical expressions may lead to human errors in the coding process, making the whole numerical implementation process even more technically challenging, highly prone to errors, and therefore takes a large amount of time.

Efforts were made to automate the implementation of hyperelastic materials. One approach is to use the computer algebra system, for example user-defined materials for ABAQUS may be automatically implemented with Mathematica [12]; but due to the limited power of existing symbolic calculation algorithms, this method is only applicable to one specific subset of hyperelastic materials: their strain energy functions need to be explicitly expressed in terms of the Lagrangian strain tensor. A more generalizable approach is to use numerical differentiation, such as obtaining the tangent modulus by perturbing the stress tensor [13,14]. While being very effective in reducing part of the workload, this method still requires the analytical derivation of the stress tensor from strain energy function, and therefore is not fully automatic.

Here, we introduce a fully automatic implementation of hyperelastic materials based on a double numerical differentiation algorithm. In addition, we searched for the optimal size of perturbation by performing a parameter sweep experiment. Implementation was done with the commercial FE platform ABAQUS, in which we verified this algorithm with both single-element models and full 3D simulation of an artery inflation test.

**METHOD**

We utilized a double numerical differentiation algorithm to achieve automation. We first perturb the strain energy function to obtain the stress tensor, and then perturb the stress tensor to obtain the tangent modulus. We then searched for the optimal perturbation size by minimizing the error between analytically calculated stress and tangent modulus values and numerical solutions. Using the resultant perturbation size, we implemented this algorithm in ABAQUS with user-defined material subroutine (UMAT) and verified through both single-element and full 3D artery inflation models. Verification results showed both high accuracy and good convergence rate.

**Mathematical Derivation**

We denote the reference and deformed configurations as $\Omega_0$ and $\Omega$ respectively, where a general mapping $\chi: \Omega_0 \to \mathbb{R}^3$ transforms a material point $X \in \Omega_0$ to $x = \chi(X, t) \in \Omega$ at time $t$. Then we can obtain deformation gradient as $F = \frac{\partial x}{\partial X}$, the right Cauchy-Green deformation tensor as $C = F^T F$, and the Lagrangian Green strain $E = \frac{1}{2}(C - I)$, where $I$ denotes the second order identity tensor. For any hyperelastic material, we have a specified strain energy function of the deformation gradient, $\Psi(F)$. Analytically, we know the 2nd Piola-Kirchhoff stress from $S = \frac{\partial \Psi}{\partial E}$, which implies its linearized form of

$$\Delta \Psi = S : \Delta E \qquad (1)$$

where the perturbation of Lagrangian Green strain can be expressed in terms of perturbation of the deformation gradient,

$$\Delta \boldsymbol{E} = \frac{1}{2}[(\boldsymbol{F}^T \Delta \boldsymbol{F})^T + (\boldsymbol{F}^T \Delta \boldsymbol{F})] \qquad (2)$$

Now, we choose the form of perturbation on $(i,j)$th component of deformation gradient to be

$$\Delta \boldsymbol{F}^{(ij)} \approx \frac{\varepsilon_S}{2} \boldsymbol{F}^{-T} (\boldsymbol{e}_i \otimes \boldsymbol{e}_j + \boldsymbol{e}_j \otimes \boldsymbol{e}_i) \qquad (3)$$

with $\{\boldsymbol{e}_i\}_{i=1,2,3}$ denoting the basis vectors and $\varepsilon_S$ denoting a small perturbation parameter (note that $i,j$ are not free indices in the Einstein summation notation). Therefore it follows the perturbation on the $(i,j)$th component of Lagrangian Green strain to be

$$\Delta \boldsymbol{E}^{(ij)} \approx \frac{\varepsilon_S}{2} (\boldsymbol{e}_i \otimes \boldsymbol{e}_j + \boldsymbol{e}_j \otimes \boldsymbol{e}_i) \qquad (4)$$

Therefore, the perturbed strain energy can be calculated as

$$\Delta \Psi^{(ij)} \approx \Psi(\boldsymbol{F}) - \Psi(\widehat{\boldsymbol{F}}^{(ij)}) \qquad (5)$$

where $\widehat{\boldsymbol{F}}^{(ij)} = \boldsymbol{F} + \Delta \boldsymbol{F}^{(ij)}$ is the perturbed deformation gradient. And also recall that

$$\Delta \Psi^{(ij)} = \boldsymbol{S} : \Delta \boldsymbol{E}^{(ij)} \approx \boldsymbol{S} : \frac{\varepsilon_S}{2} (\boldsymbol{e}_i \otimes \boldsymbol{e}_j + \boldsymbol{e}_j \otimes \boldsymbol{e}_i) = \frac{\varepsilon_S}{2} (S_{ij} + S_{ji}) \qquad (6)$$

By exploiting the symmetric properties of the 2nd Piola-Kirchhoff stress, we can eventually get the $(i, j)$th component of the 2nd Piola-Kirchhoff stress

$$S_{ij} \approx \frac{1}{\varepsilon_S} \Delta \Psi^{(ij)} = \frac{\Psi(\boldsymbol{F}) - \Psi(\widehat{\boldsymbol{F}}^{(ij)})}{\varepsilon_S} \qquad (7)$$

And with a push-forward operation we will obtain the Cauchy stress

$$\boldsymbol{\sigma} = J^{-1} \boldsymbol{F} \boldsymbol{S} \boldsymbol{F}^T \qquad (8)$$

where $J = |\boldsymbol{F}|$ denotes the Jacobian.

Applying the perturbation from Sun et al. [14], we can get the tangent modulus with respect to Jaumann objective rate to be

$$\mathbb{C}^{\sigma J} = \frac{1}{J\varepsilon_c}[\sigma(\widetilde{\boldsymbol{F}}^{(ij)}) - \sigma(\boldsymbol{F})] \qquad (9)$$

where

$$\widetilde{\boldsymbol{F}}^{(ij)} = \boldsymbol{F} + \frac{\varepsilon_c}{2}(\boldsymbol{e}_i \otimes \boldsymbol{e}_j \boldsymbol{F} + \boldsymbol{e}_j \otimes \boldsymbol{e}_i \boldsymbol{F}) \qquad (10)$$

and $\varepsilon_c$ is the small perturbation parameter.

Above concludes a sufficient implementation which uses tangent modulus in Jaumann objective rate such as ABAQUS. For other software that would require an elasticity tensor with respect to the Oldroyd rate, a conversion [15] may be done using

$$\mathbb{C}^{\sigma C}_{ijkl} = \mathbb{C}^{\sigma J}_{ijkl} - \frac{1}{2}(\delta_{ik}\sigma_{jl} + \delta_{il}\sigma_{jk} + \delta_{jk}\sigma_{il} + \delta_{jl}\sigma_{ik}) \qquad (11)$$

where $\delta_{ij}$ is the Kronecker delta. To the best of the authors' knowledge, Equation ( 3 ) - ( 7 ) and its combined usage with Equation ( 9 ) have not been reported in the literature before.

**Parameter Selection**

The selection for the perturbation parameters of $\varepsilon_s$ and $\varepsilon_c$ dictates the not only the accuracy of the numerically approximated stress, but also of the tangent modulus and therefore the convergence rate.

<u>Parameter sweep experiment set-up.</u> To choose their optimal values, we sweep the parameter space and search for the closest numerical approximation from Equation ( 7 ) - ( 9 ) compared against analytical one in Equation ( 15 ) - ( 16 ). For both $\varepsilon_s$ and $\varepsilon_c$, we evaluated a total number of 16 values from $10^{-1}$ to $10^{-16}$. We utilized three scenarios: uniaxial tension/compression, where $0.25 \leq \lambda_{11} \leq 4$; biaxial tension, where $1 \leq \lambda_{11} = \lambda_{22} \leq 4$; and simple shear, where $0 \leq \lambda_{12} \leq 0.5$. Other components of the deformation gradient were adjusted accordingly to maintain $J = 1$ in all three cases.

<u>Evaluation metrics.</u> We used the fraction of variance unexplained ($FVU = \frac{SS_{res}}{SS_{tot}}$) as the metric to evaluate the accuracy of both stress and tangent modulus. The value of the $FVU$ is equal to $1 - R^2$, where $R^2$ is the coefficient of determination that is a standard metric to describe the goodness of a model. The advantage of using $FVU$ instead of coefficient of determination is that we can closely examine the error by using the log-scale. For stress, we varied the amount of deformation and evaluated its largest principal component $\sigma_{11}$ for uniaxial and biaxial case, and shear component $\sigma_{12}$ for simple shear case:

$$FVU_{\sigma_{ij}} = \left(\frac{SS_{res}}{SS_{tot}}\right)_{\sigma_{ij}} = \frac{\sum_t \left(\sigma_{ij\,numerical}^t - \sigma_{ij\,analytical}^t\right)^2}{\sum_t \left(\sigma_{ij\,analytical}^t - \overline{\sigma_{ij\,analytical}}\right)^2} \quad (12)$$

where $t$ denotes the superscript for different deformation levels. For tangent modulus, we evaluated all of its components at only the largest deformation in all three cases:

$$FVU_{\mathbb{C}^{\sigma J}} = \left(\frac{SS_{res}}{SS_{tot}}\right)_{\mathbb{C}^{\sigma J}} = \frac{\sum_{i,j,k,l}\left(\mathbb{C}_{ijkl\,numerical}^{\sigma J} - \mathbb{C}_{ijkl\,analytical}^{\sigma J}\right)^2}{\sum_{i,j,k,l}\left(\mathbb{C}_{ijkl\,analytical}^{\sigma J} - \overline{\mathbb{C}_{analytical}^{\sigma J}}\right)^2} \quad (13)$$

<u>Material constitutive model.</u> Neo-Hookean model was used in this step, the strain energy function of which is defined as

$$\Psi = C_{10}(\bar{I}_1 - 3) + \frac{1}{D}(J - 1)^2 \quad (14)$$

where $C_{10}$ and $D$ are the material constants characterizing the isochoric and volumetric responses of the material, and take the values of $C_{10} = 80 \times 10^3 \text{ Pa}, D = 2 \times 10^{-6} \text{ Pa}^{-1}$; $\bar{I}_1 = \text{tr}(\bar{\boldsymbol{C}})$ denotes the first invariant of the deviatoric right Cauchy-Green tensor $\bar{\boldsymbol{C}} = J^{-\frac{2}{3}}\boldsymbol{C}$. Its well-known analytical expression of stress and tangent modulus are

$$\boldsymbol{\sigma} = \frac{2C_{10}}{J}\left(\bar{\boldsymbol{b}} - \frac{1}{3}tr(\bar{\boldsymbol{b}})\boldsymbol{I}\right) + \frac{2}{D}(J-1)\boldsymbol{I} \quad (15)$$

$$\mathbb{C}^{\sigma J}_{ijkl} = \frac{2C_{10}}{J}\left[\frac{1}{2}(\delta_{ik}\bar{b}_{jl} + \bar{b}_{ik}\delta_{jl} + \delta_{il}\bar{b}_{jk} + \bar{b}_{il}\delta_{jk}) - \frac{2}{3}(\delta_{ij}\bar{b}_{kl} + \bar{b}_{ij}\delta_{kl}) + \frac{2}{9}\delta_{ij}\delta_{kl}\bar{b}_{mm}\right] + \frac{2}{D}(2J-1)\delta_{ij}\delta_{kl} \quad (16)$$

where $\bar{\boldsymbol{b}} = J^{-\frac{2}{3}}\boldsymbol{b} = J^{-\frac{2}{3}}\boldsymbol{F}\boldsymbol{F}^T$ is the deviatoric left Cauchy-Green tensor.

**Single-element Verification**

After identifying the optimal perturbation parameters, we performed single-element FE analysis to verify this algorithm.

<u>FE model set-up.</u> To be consistent with the previous parameter selection experiments, we also tested three cases of uniaxial tension/compression, biaxial tension and simple shear. Following the previous study [14], three-dimensional brick element with reduced integration (C3D20R in ABAQUS) was used for the single-element model in ABAQUS Standard. The dimensions of the cube is $1 \times 1 \times 1$ mm. 0.75 mm uniaxial compression load, 3 mm uniaxial extension load, 3 mm equi-biaxial extension load, and 1 mm simple shear load were prescribed as boundary conditions in 20 fixed and equal sized increments.

<u>Material constitutive model.</u> The same Neo-Hookean solids with $C_{10} = 80 \times 10^3 \text{ Pa}, D = 2 \times 10^{-6} \text{ Pa}^{-1}$ were also used. We compared two implementations of the material constitutive laws: the analytical built-in model of ABAQUS, and the numerical model using user defined material subroutine (UMAT) for Equation ( 7 ) - ( 9 ). Perturbation parameters for stress and tangent modulus were chosen as $\varepsilon_s = 10^{-6}$ and $\varepsilon_c = 10^{-4}$, which is the optimal result in the parameter selection experiments.

**Full 3D Model Verification**

Finally, following prior work [14,16], we applied our numerical implementation on a more sophisticated aorta inflation model with an anisotropic hyperelastic solid with two families of fibers.

<u>FE model set-up.</u> To represent a segment of rat artery, we used a quarter-symmetric model (Figure 1) with 450 3D continuum brick elements (C3D8). The model contains 6 layers of elements on the radial direction through the wall, 3 layers on the axial direction of the artery and 25 layers on the 90° section of the circumference. The symmetric axes were applied fixed boundary condition on all x, y and z degrees of freedom. The pressure load of 25 kPa was applied on the inner walls of the artery, and automatic increment control was used with total step time of 1 sec, minimum increment size of $1 \times 10^{-5}$ sec and maximum of 1 sec .

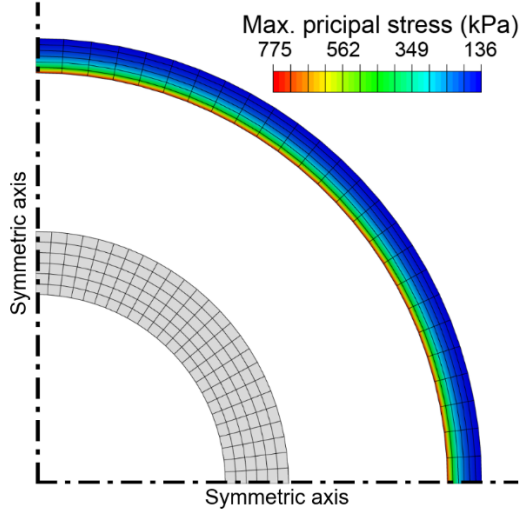

*Figure 1 Screenshot of the full 3D verification simulating the artery inflation with a quarter-symmetric FE model. The grey mesh denotes the undeformed configuration of the artery section, and they rainbow-colored mesh denotes the principal stress distribution in the deformed configuration at* $25$ kPa *pressure load.*

Material constitutive model. The strain energy function and the stress for the Holzapfel model, describing transversely isotropic solids with two families of fibers, is

$$\Psi = C_{10}(\bar{I}_1 - 3) + \frac{k_1}{2k_2}\left[(e^{k_2(\bar{I}_4-1)^2} - 1) + (e^{k_2(\bar{I}_6-1)^2} - 1)\right] + \frac{1}{D}\left(\frac{J^2-1}{2} - \ln J\right) \quad (17)$$

where $\bar{I}_4$ and $\bar{I}_6$ are strain pseudo-invariants of $\bar{\mathbf{C}}$ and equal the squares of stretches in each of the fiber directions. They are defined as $\bar{I}_4 = \boldsymbol{a_0} \cdot \bar{\mathbf{C}}\boldsymbol{a_0}$ and $\bar{I}_6 = \boldsymbol{g_0} \cdot \bar{\mathbf{C}}\boldsymbol{g_0}$ each, where $\boldsymbol{a_0}$ and $\boldsymbol{g_0}$ are the fiber orientation vectors in the reference configuration. $C_{10}, D, k_1, k_2$ are material constants and take the values of $C_{10} = 2.212 \times 10^4$ Pa, $D = 1 \times 10^{-6}$ Pa$^{-1}$, $k_1 = 206$ Pa, $k_2 = 1.465$. The preferred collagen fiber orientations are $\pm 39.76°$ with respect to the artery circumferential direction. We compared the analytical implementation of this model (built-in ABAQUS material) with the numerical implementation presented in this study. For the latter, perturbation parameters for stress and tangent modulus were also chosen as $\varepsilon_S = 10^{-6}$ and $\varepsilon_C = 10^{-4}$ according to the parameter selection experiments. The analytic solutions for stress and tangent modulus of the Holzapfel model are not needed to implement the presented numerical algorithm and is therefore not shown; interested readers may refer to Gasser et al. [2] for the comprehensive derivation.

**RESULTS**

Parameter selection. We compared the numerically estimated stress with the accurate analytical solution using $\varepsilon_S = 10^{-6}$ and $10^{-8}$, and both perturbation magnitudes showed small error (Figure 2A-C). When we choose the threshold $FVU$ as $10^{-4}$, a comprehensive exploration of the parameter shows that high accuracies in stress were achieved for all $10^{-4} \leq \varepsilon_S \leq 10^{-12}$ and all three cases of uniaxial compression/tension, biaxial tension and simple shear (Figure 2D-F). The optimal perturbation magnitude to obtain stress is $10^{-8}$. For the accuracy for tangent modulus, we also found a range of combinations of $\varepsilon_S$ and $\varepsilon_C$ achieving $FVU$ smaller than $10^{-4}$, and the optimal combination is $\varepsilon_S = 10^{-6}$, $\varepsilon_C = 10^{-4}$ (Figure 2G-I). We noticed that this $\varepsilon_S$ is different from the optimal value of $10^{-8}$ above in terms of stress accuracy. However, in that case the error on tangent modulus is slightly larger (Figure 2G-I) due to the round-off error after cascading from numerical perturbation for stress. We eventually took a compromise, so the tangent modulus is at its best accuracy while the stress also has high resolution with $FVU$ smaller than $10^{-10}$.

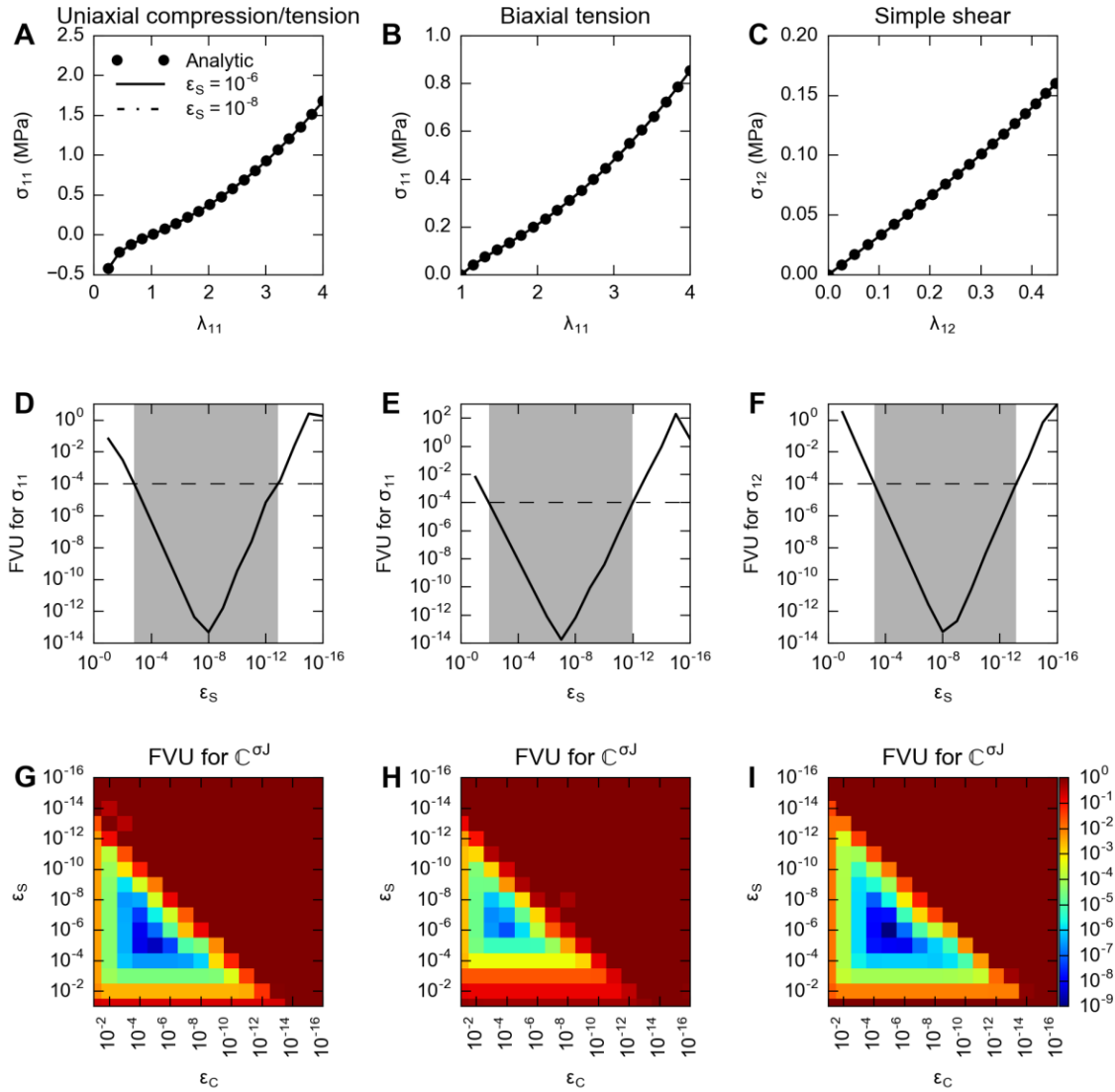

*Figure 2 Results for the parameter selection experiment. Three columns each represent cases of uniaxial compression/tension, biaxial tension and simple shear. A-C: comparison of the stress-stretch curves of analytic solution, $\varepsilon_s = 10^{-6}$ and $10^{-8}$. D-F: FVU for stresses, plotted in common log scale. G-I: FVU for the tangent modulus, plotted in common log scale.*

<u>Single-element verification.</u> The convergence rate and the stress magnitude from the analytical and numerical material implementations were comparable in all model runs with uniaxial, biaxial and simple shear loading conditions. The total number of iterations as an indicator of convergence rate, and the relative error in stress as an indicator of accuracy are shown in Table 1. Average relative error for the four loading conditions is $7.31 \times 10^{-5}$.

Table 1 The total number of iterations and the relative error in stress for single-element verification.

| Uniaxial compression | | | Uniaxial tension | | |
|---|---|---|---|---|---|
| Numerical | Analytic | | Numerical | Analytic | |
| Iter # | Iter # | Stress rel err | Iter # | Iter # | Stress rel err |
| 22 | 21 | 1.69E-04 | 27 | 25 | 5.20E-05 |
| Biaxial tension | | | Simple shear | | |
| Numerical | Analytic | | Numerical | Analytic | |
| Iter # | Iter # | Stress rel err | Iter # | Iter # | Stress rel err |
| 41 | 41 | 5.54E-05 | 21 | 21 | 1.58E-05 |

Full 3D model verification. Artery inflation simulation also showed comparable performance of convergence and accuracy between the analytical and numerical implementations. Both agreed well with the experimentally measured outer radius (Figure 3A), and have similar performance in terms of total number of steps attempted, number of iterations and predicted artery outer radii as shown in Table 2. At maximum pressure of $25 \text{ kPa}$, the relative error in predicted radius is $4.25 \times 10^{-6}$. The underformed and deformed view of the transmural stress distribution was already shown in Figure 1. To closely examine this distribution under different pressure levels, we have plotted principal stresses under 6 equal increment pressure loads ($0, 6.25, 12.5, 18.75$ and $25 \text{ kPa}$) using linear interpolation, with respect to normalized radius where 0 represent the inner and 1 the outer wall of the artery (Figure 3B). With an increase in pressure, the stress distribution becomes less uniform and we observe higher stress near the inner wall of the artery. The analytical and numerical material implementation yield the same result.

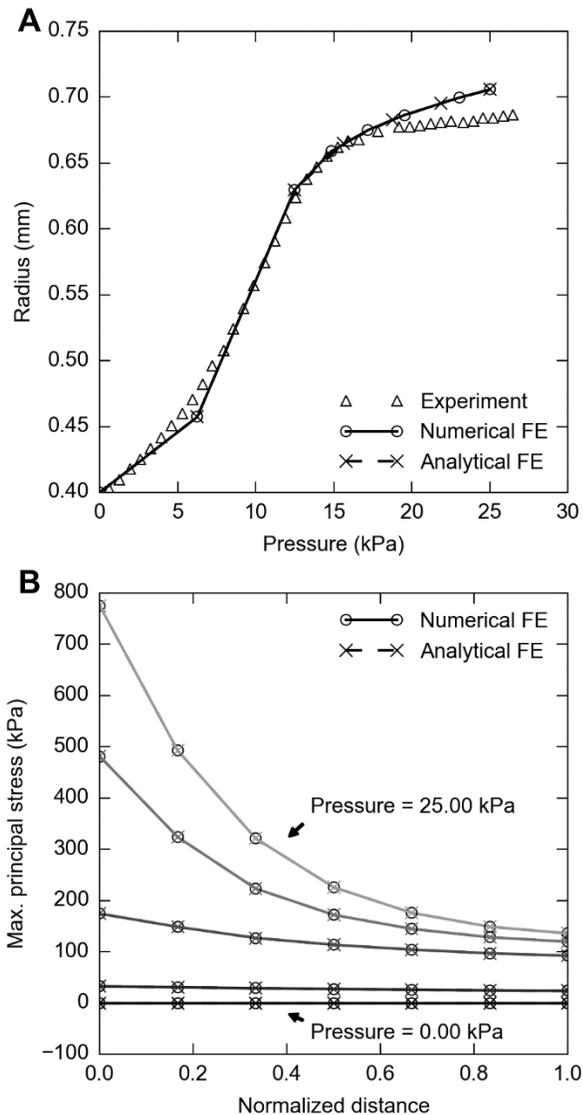

*Figure 3 Result of the full anisotropic model verification for the artery inflation. A: Comparison of the radius-pressure curve for experiment, FE implementation with analytical solution and numerical solution. B: Comparison of the max. principal stress vs. normalized distance curve for FE implementation with analytical solution and numerical solution. Line color from dark to light each represent pressure load of $0, 6.25, 12.5, 18.75$ and $25$ kPa. With an increase in pressure, the stress distribution becomes less uniform and*

*we observe higher stress near the inner wall of the artery. Both analytical and numerical material implementation yield the same result.*

Table 2 *The total number of steps attempted, number of iterations and predicted artery outer radii were compared between the analytical and numerical implementations for the full 3D model verification simulating artery inflation.*

| | Numerical | | | | | Analytic | | | |
|---|---|---|---|---|---|---|---|---|---|
| Inc # | Att # | Iter # | Pressure (kPa) | Radius (mm) | Inc # | Att # | Iter # | Pressure (kPa) | Radius (mm) |
| 1 | 1U | 1 | | | 1 | 1U | 1 | | |
| 1 | 2 | 4 | 6.25 | 0.46 | 1 | 2 | 4 | 6.25 | 0.46 |
| 2 | 1 | 4 | 12.50 | 0.63 | 2 | 1 | 8 | 12.50 | 0.63 |
| 3 | 1U | 3 | | | 3 | 1U | 9 | | |
| 3 | 2 | 6 | 14.84 | 0.66 | 3 | 2 | 9 | 15.63 | 0.67 |
| 4 | 1 | 4 | 17.19 | 0.67 | 4 | 1 | 4 | 18.75 | 0.68 |
| 5 | 1 | 3 | 19.53 | 0.69 | 5 | 1 | 3 | 21.88 | 0.70 |
| 6 | 1 | 3 | 23.05 | 0.70 | 6 | 1 | 2 | 25.00 | 0.71 |
| 7 | 1 | 2 | 25.00 | 0.71 | | | | | |
| Total | 9 | 30 | | | | 8 | 40 | | |

**DISCUSSION**

In this study, we presented the first fully automated method to implement arbitrary hyperelastic material for finite element analysis, where only the strain energy function needs to be defined. This numerical method has good accuracy and convergence performance comparable with the traditional analytical methods, as shown by the verification experiments.

We would like to note that this algorithm is not a complete replacement for the traditional analytical implementation of hyperelastic materials. In the trade-off between human and CPU time, this algorithm largely reduced the former at the cost of an increase in the latter. Although in the single-element and full 3D verification tests, the number of iterations towards convergence for both numerical and analytical algorithms are comparable, the cost for material evaluation in each iteration is different. Even when we fully exploit the symmetry properties, the numerical algorithms needs to perform at least 6 perturbations to obtain all 6 independent stress components, and also 6 perturbations for each the 21 independent tangent modulus components. Compared with the direct evaluation of the analytical algorithms, the algorithm presented within is more computationally expensive, which will be an issue especially for large-scale models.

Nevertheless, while this automatic algorithm saves the more expensive human time, the additional computational cost is trivial for small-scale models. Take the artery inflation experiment as an example, a computer with 8-core 3.4 GHz CPU and 16 GB of memory takes 4.2 seconds to run with the numerical implementation, while the analytical one takes 3.8 seconds. As a significant amount of time is often on testing different material models or verifying the material implementation during development, using the numerical implementation presented herein will tremendously accelerate the process.

For large scale models, while the analytical implementation should be used to reduce the computational cost, the numerical algorithm presented will be a great debugging tool. Since only the strain energy function is required, this algorithm leaves little space for human error and always yield correct stress and tangent modulus. During the analytical implementation, it was difficult to know whether the derived formula were correct; by comparing the derivation result and numerical solution, one can quickly identify human errors made.

In the future, we plan to explore further to expand the usage of the presented double differentiation algorithm. For very stiff hyperelastic strain energy functions, the numerical differentiation will have limited accuracy, which we may solve with adaptive perturbation size and arbitrary-precision arithmetic. In addition, the pure numerical algorithm opens the door towards novel hyperelastic models that may not have an analytical expression for stress and tangent modulus.

**ACKNOWLEDGMENT**

We thank Dr. Silvia Blemker for the inspiring discussions on this study, and Ms. Lingtian Wan for help in the manuscript preparation.

**FUNDING**

This work was supported by a grant from the National Institutes of Health (NINDS R01NS073119 to EAL and GJG). The content is solely the responsibility of the authors and does not necessarily represent the official views of the U.S. National Institutes of Health (NIH).